# On the implementation of dislocation reactions in continuum dislocation dynamics modeling of mesoscale plasticity

Vignesh Vivekanandan[1], Peng Lin[1], Grethe Winther[2], Anter El-Azab[1]

[1] School of Materials Engineering, Purdue University, Neil Armstrong Hall of Engineering, 701 W Stadium Avenue, West Lafayette, IN 47907, USA

[2] Department of Mechanical Engineering, Technical University of Denmark, DK-2800 Kgs. Lyngby, Denmark

Abstract. The continuum dislocation dynamics framework for mesoscale plasticity is intended to capture the dislocation density evolution and the deformation of crystals when subjected to mechanical loading. It does so by solving a set of transport equations for dislocations concurrently with crystal mechanics equations, with the latter being cast in the form of an eigenstrain problem. Incorporating dislocation reactions in the dislocation transport equations is essential for making such continuum dislocation dynamics predictive. A formulation is proposed to incorporate dislocation reactions in the transport equations of the vector density-based continuum dislocation dynamics. This formulation aims to rigorously enforce dislocation line continuity using the concept of virtual dislocations that close all dislocation loops involved in cross slip, annihilation, and glissile and sessile junction reactions. The addition of virtual dislocations enables us to accurately enforce the divergence free condition upon the numerical solution of the dislocation transport equations for all slip systems individually. A set of tests were performed to illustrate the accuracy of the formulation and the solution of the transport equations within the vector density-based continuum dislocation dynamics. Comparing the results from these tests with an earlier approach in which the divergence free constraint was enforced on the total dislocation density tensor or the sum of two densities when only cross slip is considered shows that the new approach yields highly accurate results. Bulk simulations were performed for a face centered cubic crystal based on the new formulation and the results were compared with discrete dislocation dynamics predictions of the same. The microstructural features obtained from continuum dislocation dynamics were also analyzed with reference to relevant experimental observations.

# 1. Introduction

It is currently well established that the phenomenological continuum plasticity theories are insufficient to capture the plastic behavior of crystals at the scale of the dislocation microstructure. This motivated the metal deformation community to develop plasticity models based upon the dislocation mechanics. The method of discrete dislocation dynamics (DDD) simulation was developed as a part of that effort, and, in spite of its computational limitations, the method has enabled many successful mesoscale plasticity investigations (Arsenlis et al., 2007; Groh and Zbib, 2009; Kubin et al., 1992; Po et al., 2014; Rhee et al., 1998; Weygand et al., 2002). Relevant to this current work is the use of DDD models in understanding the role of junctions in plastic deformation. Indeed, the success of these models inspired the study of the role of dislocation junctions in stage II strain hardening behavior (Franciosi et al., 1980; Kocks, 1966; Taylor, 1934). Dislocation junctions in face centered cubic (FCC) crystals are categorized into four categories: Hirth junctions, Lomer junctions, glissile junctions and collinear junctions. The role of such junctions in the strain hardening behavior of crystals was recently studied by many discrete dislocation dynamics simulations (Devincre et al., 2008; Devincre, 2013; Stricker and Weygand, 2015; Sills et al., 2018; Stricker et al., 2018; Mishra and Alankar, 2019)). It was found that glissile junctions have the highest contribution towards hardening (Sills et al., 2018). Apart from its contribution to hardening, glissile junctions were also found to be responsible for the dislocation multiplication mechanism (Stricker et al., 2018) in FCC metals, which play a significant role in influencing microstructure evolution. DDD simulations were also used to study cross slip, which is a thermally activated process that results in the change of the glide plane of the screw-orientated dislocations. Similar to glissile junctions, cross slip also contributes towards the dislocation multiplication mechanism playing a crucial role in the microstructure evolution (Hussein et al., 2015; Stricker et al., 2018). Although DDD simulations have demonstrated sufficient capability to study dislocation behavior under deformation, they suffer from scalability issues making them unsuitable for large-scale bulk simulations, and this is where continuum dislocation dynamics (CDD) is thought to fill an important gap.

Early attempts to relate plastic deformation and dislocations at a continuum level resulted in the now-called classical continuum theory of dislocations (Kröner, 1959; Nye, 1953). In these classical works, the incompatibility of the elastic and plastic distortion field was cast in the form of a continuous field called the dislocation density tensor. However, the lack of sufficient information to determine the dislocation motion from the dislocation density tensor, hindered further development of this theory toward being a plasticity framework (Kosevich, 1965; Mura, 1963). Inspired by Mura and Kosevich, Acharya and co-workers (Acharya and Roy, 2006; Roy and Acharya, 2006) developed a model called field dislocation mechanics, in which transport laws for the evolution of the dislocation density tensor along with a closure law for the speed of dislocation transport in terms of the local stress were used to build a plasticity framework based on a tensor representation of the dislocation field. Several other groups, however, used statistical approaches to reach continuum, density-based descriptions of dislocation dynamics in two-dimensions (2D) (Groma, 1997; Groma et al., 2003; Kooiman et al., 2014; Zaiser et al., 2001). These 2D models were later used to capture the evolution of the dislocation density in a crystal plasticity type setting by (Yefimov et al., 2004). Extension of the statistical approach to three dimensions (3D) was then accomplished by El-Azab and co-workers (El-Azab, 2000a, 2006; Deng and El-Azab, 2009), where the curved interconnected dislocation configurations were represented in a phase space following the classical statistical mechanics concepts. Other different approaches were pursued to capture the 3D dislocation microstructure. For example, Hochrainer and co-workers developed a class of models (Hochrainer et al., 2007; Sandfeld et al., 2010) that represent dislocations using a tensor in a higher dimensional phase space which carries the additional information about the local line orientation and curvature of the curved lines. A simplified version of the higher-order theory suitable for numerical implementation was developed later (Hochrainer, 2015; Hochrainer et al., 2014; Sandfeld et al., 2011; Sandfeld and Zaiser, 2015) based on the concept of dislocation alignment tensors (Hochrainer, 2015).

The first report of modeling of cross slip and dislocation reactions in CDD was due to El-Azab (El-Azab, 2000a), where slip system-level consideration of dislocations along with their line directions enabled the introduction of rate terms representing such processes into the dislocation transport equations; see also (Deng and El-Azab, 2009). In (Deng and El-Azab, 2010), time series analysis was introduced to estimate the cross slip and junction reaction rates from DDD data. More recently, Monavari and Zaiser (Monavari and Zaiser, 2018) proposed a model that incorporates dislocation reactions like cross slip and glissile junctions as sources of dislocation multiplication. Their model associated dislocation reactions with a new field variable called curvature density, which described the volume density of dislocation loops in the system. Recently, Sudmanns and co-workers (Sudmanns et al., 2019) also developed a model along the similar lines in which they represented dislocation reactions based on the lessons learned from DDD simulations (Stricker et al., 2018). The dislocation density evolution on the active and inactive slip systems due to glissile junction reaction was captured successfully similar to the results obtained from DDD simulations.

In this paper, a new formulation to incorporate dislocation cross slip and reactions in CDD is proposed and implemented within the framework previously developed in (Xia and El-Azab, 2015; Lin and El-Azab, 2020; and Starkey et al., 2020). In this framework, dislocations are represented by vector density fields on individual slip systems that appears locally as line bundles with a single line direction at each point in space. This bundle representation of dislocations is motivated by the need to obtain density field solutions at the scale of several tens of nanometers in order to resolve the density variations of highly heterogeneous dislocation structures. At such a high resolution, dislocations much have single line directions at each point in space. The evolution of the dislocation density field is governed by a curl type dislocation transport equation with dislocation reaction terms. The current formulation aims to represent the change in the slip system density due to dislocation cross slip and reactions using a closure density called here the virtual dislocation density. Dislocation processes like cross slip and glissile junctions are modeled as dislocation source terms in the transport equations. In addition to that, a new method of enforcing the divergence-free constraint on the dislocation density is introduced that takes virtual density into consideration, thus enabling the implementation of this important constraint to individual slip system densities. The divergence-free constraint itself is enforced upon the numerical solution of the dislocation transport equations to guarantee that the numerical errors, however small they might be, do not propagate into the solution and as a result obliterate its physical nature. This significantly facilitates the computational solution of the dislocation transport system of equations in 3D, and enables us to track the dislocation network effectively. The paper is organized as follows. Section 2 outlines the CDD formalism of mesoscale plasticity used here. Section 3 introduces the concept of virtual density and how it is integrated into the dislocation kinetics. Section 4 describes the numerical implementation of the dislocation kinetics equations. Section 5 discusses the simulation results obtained based on this formulation, followed by a short discussion and conclusions.

## 2. Continuum dislocation dynamics model

Continuum dislocation dynamics formulation of mesoscale plasticity comprises of two parts, namely the stress-equilibrium problem and dislocation kinetics problem. In this section, the method used to solve the stress equilibrium problem is outlined and the dislocation kinetics problem is discussed in detail.

### 2.1 Stress-equilibrium problem

For the case of small deformation, the distortion tensor, $\boldsymbol{\beta}$, can be expressed as the gradient of displacement, $\boldsymbol{\nabla u}$, which can be decomposed into two incompatible fields, the elastic distortion tensor, $\boldsymbol{\beta}^e$, and plastic distortion tensor, $\boldsymbol{\beta}^p$. These two tensors describe the deformation of the lattice and the shape change of the crystal, respectively. The displacement field due to the applied load and dislocations present in the crystal is obtained by solving the eigen distortion problem stated below:

$$\boldsymbol{\nabla} \cdot \boldsymbol{\sigma} \;=\; \boldsymbol{\nabla} \cdot (\boldsymbol{C}\!:\!\boldsymbol{\beta}^e) = \boldsymbol{\nabla} \cdot \bigl(\boldsymbol{C}\!:\!(\boldsymbol{\nabla u} - \boldsymbol{\beta}^p)\bigr) \;=\; \boldsymbol{0} \text{ in } \Omega \qquad (1)$$
$$\boldsymbol{u} = \bar{\boldsymbol{u}} \text{ on } \partial\Omega_u$$
$$\boldsymbol{n} \cdot \boldsymbol{\sigma} \;=\; \bar{\boldsymbol{t}} \text{ on } \partial\Omega_t.$$

In the above, $\boldsymbol{\sigma}$ is Cauchy stress tensor, $\boldsymbol{C}$ the elastic tensor, $\Omega$ the domain of solution, $\partial\Omega_u$ and $\partial\Omega_t$, respectively, are the parts of the boundary on which the displacement $\boldsymbol{u} = \bar{\boldsymbol{u}}$ and the traction $\bar{\boldsymbol{t}}$ are prescribed. By solving the above boundary value problem, the displacement and stress fields are updated, and the latter is then used to determine the Peach-Koehler force and the dislocation velocity based on a mobility law. Together with the dislocation density, the dislocation velocity is used to find the rate of plastic distortion via Orowan's law. The latter is then used to update the plastic distortion using a direct time integration (Xia and El-Azab, 2015) or via the field dislocation mechanics scheme (Acharya and Roy, 2006; Roy and Acharya, 2005, Lin et al., 2020).

## 2.2 Dislocation kinetics

In continuum dislocation dynamics, dislocations are represented as continuous fields in terms of dislocation density tensor, $\boldsymbol{\alpha}$. In the current work, the tensor $\boldsymbol{\alpha}$ is defined following the formulation proposed by (Kröner, 1959) where it is quantified in terms of plastic distortion tensor $\boldsymbol{\beta}^{\mathrm{p}}$ as

$$\boldsymbol{\alpha} = -\nabla \times \boldsymbol{\beta}^{\mathrm{p}}, \qquad (2)$$

The evolution of $\boldsymbol{\alpha}$ is obtained by differentiating eq. (2) with time, which yields us

$$\dot{\boldsymbol{\alpha}} = -\nabla \times \dot{\boldsymbol{\beta}}^{\mathrm{p}}. \qquad (3)$$

In the case of single slip, and assuming that the dislocation field is characterized by a single line direction and hence a unique velocity field at every point in space, the rate of plastic distortion is obtained from Orowan's law:

$$\dot{\boldsymbol{\beta}}^{\mathrm{p}} = -\boldsymbol{v} \times \boldsymbol{\alpha}, \qquad (4)$$

with $\boldsymbol{v}$ being the dislocation velocity. Substituting this in eq. (3), we get

$$\dot{\boldsymbol{\alpha}} = \nabla \times (\boldsymbol{v} \times \boldsymbol{\alpha}). \qquad (5)$$

The unique line and velocity of the dislocation field at every point in space is known as the dislocation bundle idealization, and it assumes that the spatial resolution is fine enough so that dislocations of a given Burgers vector and slip plane have the same line direction at ever point in the crystal. This allows us to define the vector dislocation density $\boldsymbol{\rho}$ at every point as

$$\boldsymbol{\rho} = \rho \boldsymbol{\xi}, \qquad (6)$$

where $\rho$ is the scalar dislocation density and $\xi$ is the unit tangent of the dislocations. The dislocation density tensor can then be expressed in terms of dislocation density vector and Burgers vector,

$$\boldsymbol{\alpha} = \boldsymbol{\rho} \otimes \boldsymbol{b}. \tag{7}$$

Substituting eqn. (7) in eqn. (5), we get

$$\dot{\boldsymbol{\rho}} \otimes \boldsymbol{b} = \nabla \times \left( \boldsymbol{v} \times (\boldsymbol{\rho} \otimes \boldsymbol{b}) \right). \tag{8}$$

The above equation can then be simplified to

$$\dot{\boldsymbol{\rho}} = \nabla \times (\boldsymbol{v} \times \boldsymbol{\rho}). \tag{9}$$

Equation (9) is referred to as the dislocation transport equation, and it accounts for the increase in line length of the dislocations due to curvature. As explained in Section 5.4, the dislocation transport velocity $\boldsymbol{v}$ is estimated in terms of the local resolved shear stress by using a mobility law. The above equation is incomplete since dislocation transport is not the only mechanism through which dislocation density evolves. Many DDD simulations have pointed out that dislocation processes such as cross slip and junction formation reactions are responsible for re-distributing dislocations amongst different slip systems (Arsenlis et al., 2007; Hussein et al., 2015; Kubin et al., 2008; Sills et al., 2018; Stricker et al., 2018; Stricker and Weygand, 2015). Hence, to complete the kinetic equations, a general formulation to incorporate dislocation reactions is proposed in the next section.

## 2.3 Dislocation reactions

In this current work we refer to cross slip, junction formation and annihilation processes as dislocation reactions. Since, the dislocations are represented as vector densities, the annihilation reaction on the same slip system is considered implicitly. Hence, in this section a method to incorporate cross slip and junction reactions into dislocation density evolution equation is described.

Cross slip is a thermally activated process through which a screw dislocation changes its glide plane to the cross-slip plane that shares the same Burgers vector. Junction formation is a process through which two dislocation segments from two different slip systems react with each other and form a third (product) segment which lies along the intersection of two planes of the reacting segments and has Burgers vector equaling to the sum of the Burgers vectors of two reacting segments. They can be classified into four types of junctions namely glissile, Lomer, Hirth, and collinear. Cross slip and glissile junction acts as dislocation sources since they introduce new dislocations into a given slip system. Unlike cross-slip and glissile junction, the other junctions do not introduce new dislocations to other slip systems, but they can be unzipped and their content is reintroduced into their parent slip systems.

In CDD models, dislocation reactions are characterized as rate processes that captures the fraction of dislocation density that is consumed from or added to a given slip system density. In FCC crystals, dislocations in a slip system can be involved in at most two cross slip and 11 junction reactions. The rates of these processes are expressed as follows:

$$\dot{\rho}_{\text{cs},i \to j} = \dot{r}_{i \to j} \rho_g^i \tag{10}$$

$$\dot{\rho}_{\text{G},ijk} = \dot{r}_{\text{G},ijk} (\rho_g^i \cdot e_{ij})(\rho_g^j \cdot e_{ij}) e_{ij} \tag{11}$$

$$\dot{\rho}_{\text{L},ij} = \dot{r}_{\text{L},ij} (\rho_g^i \cdot e_{ij})(\rho_g^j \cdot e_{ij}) e_{ij} \tag{12}$$

$$\dot{\rho}_{\text{H},ij} = \dot{r}_{\text{H},ij} (\rho_g^i \cdot e_{ij})(\rho_g^j \cdot e_{ij}) e_{ij}, \tag{13}$$

where $\dot{\rho}_{\text{cs},i \to j}, \dot{\rho}_{\text{G},ijk}, \dot{\rho}_{\text{L},ij}, \dot{\rho}_{\text{H},ij}$ represent the rate of change of dislocation density due to cross slip and glissile, Lomer and Hirth junctions, respectively, $\dot{r}_{i \to j}, \dot{r}_{\text{G},ijk}, \dot{r}_{\text{L},ij}$ and $\dot{r}_{\text{H},ij}$ are the corresponding rate coefficients, $\rho_g^i$ denotes the glide dislocation density of slip system $i$, and $e_{ij}$ is the unit vector along the line of intersection of the two slip planes $i$ and $j$ involved in the reaction. In the case of cross slip, the subscript $i \to j$ indicates that the cross slip event happens from slip system $i$ to slip system $j$. In the case of glissile junctions, the indices $i, j$ and $k$ refer to the dislocations on slip system $i$ and $j$ reacting to form glissile junction on slip system $k$. In the case of Lomer and Hirth locks, the indices $i, j$ refer to the dislocations on slip system $i$ and $j$ reacting to form Lomer and Hirth junctions. The rate coefficients for the different reactions are expressed in the form:

$$\dot{r}_{i \to j} = \iota_s \dot{p}_{i \to j} \tag{14}$$

$$\dot{r}_{\text{G},ijk} = \iota_{e_{ij}} \dot{p}_{\text{G},ijk} \tag{15}$$

$$\dot{r}_{\text{L},ij} = \iota_{e_{ij}} \dot{p}_{\text{L},ij} \tag{16}$$

$$\dot{r}_{\text{H},ij} = \iota_{e_{ij}} \dot{p}_{\text{H},ij}, \tag{17}$$

where $\dot{p}_{i \to j}, \dot{p}_{\text{G},ijk}, \dot{p}_{\text{L},ij}$ and $\dot{p}_{\text{H},ij}$ are the probability rates of cross slip and glissile, Lomer and Hirth junctions, respectively, all computed from discrete dislocation line models. In equation (14), $\iota_s$ is a dislocation line orientation indicator function that is non-zero only when the line orientation satisfies a screw direction criterion. It is important to note that the cross-slip can happen only if the local resolved shear stress on the cross-slip plane is greater than that of the original glide plane. In equations. (15-17), $\iota_{e_{ij}}$ is an orientation indicator function that is non-zero only when the line orientations of the reacting segments satisfy the criterion for junction formation (Lin and El-Azab, 2020). The probability rates of cross-slip and junction reactions can be estimated based on the procedure outlined by (Deng and El-Azab, 2010; Xia et al., 2016) where the statistics of these processes are obtained from DDD simulations in the form of time series and subsequently coarse grained for use in CDD simulations.

In addition to the different types of reactions mentioned above, the Lomer and Hirth junctions can be unzipped thus reintroducing dislocations into the glide density. The rate of change of the glide dislocation density in slip system $i$ due to the unzipping of Lomer and Hirth junctions is denoted here by $\dot{\boldsymbol{\rho}}_{L^*,i}$ and $\dot{\boldsymbol{\rho}}_{H^*,i}$, respectively. These rates can then be expressed as a fraction of existing sessile dislocation density that can be unzipped over a short time step (Sudmanns et al., 2020). As such, the rate of unzipping of Lomer junctions can be represented in the form $\dot{\boldsymbol{\rho}}_{L^*,i} = \dot{r}_{L^*,i}\,\boldsymbol{\rho}_{L,i}$ where $\dot{r}_{L^*,i}$ corresponds to the rate coefficient of the Lomer unzipping process in slip system $i$ and $\boldsymbol{\rho}_{L,i}$ corresponds to the Lomer density that can be unzipped back into slip system $i$. Based on the equations (10) through (13), equation (9) is closed by incorporating the source terms into the dislocation density evolution as follows

$$\dot{\boldsymbol{\rho}}_g^i = \nabla \times (\boldsymbol{v}^i \times \boldsymbol{\rho}_g^i) - \dot{\boldsymbol{\rho}}_{cs,i\to j} - \sum_j \dot{\boldsymbol{\rho}}_{G,ijk} - \sum_j \dot{\boldsymbol{\rho}}_{L,ij} - \sum_j \dot{\boldsymbol{\rho}}_{H,ij} + \dot{\boldsymbol{\rho}}_{cs,j\to i} \qquad (18)$$
$$+ \sum_{j,k} \dot{\boldsymbol{\rho}}_{G,jki} + \dot{\boldsymbol{\rho}}_{L^*,i} + \dot{\boldsymbol{\rho}}_{H^*,i}$$

The four terms following the curl term to the right-hand side of equation (18) correspond to the rate at which dislocation density is removed from the slip system $i$ due to one cross slip, four glissile junction reactions, two Lomer lock reactions and two Hirth lock reaction, respectively. The summation over $j$ in the third term refers to the slip systems $j$ with which slip system $i$ reacts to form a glissile junction on slip system $k$. The summation over $j$ for the fourth and fifth terms refers to the different slip systems $j$ with which slip system $i$ reacts to form Lomer and Hirth locks. The last four terms in the same equation correspond to the rate at which the dislocation density is added to the slip system $i$ due to one cross slip reaction, two glissile junction reactions and Lomer and Hirth locks unzipping, respectively. The summation over $j$ and $k$ in one of the terms refers to different distinct pairs of $j$ and $k$ that react to form glissile junction product on slip system $i$. In this notation, the pair $jk$ is the same as the pair $kj$, and that must not be double counted.

Finally, to ensure that dislocations do not end inside the crystal a divergence-free condition is enforced on the dislocation density tensor $\boldsymbol{\alpha}$, which can be written as

$$\nabla \cdot \boldsymbol{\alpha} = \boldsymbol{0}. \qquad (19)$$

In the case of a single slip system and in the absence of reactions, this condition reads: $\nabla \cdot \boldsymbol{\rho} = 0$. In such cases, the dislocation density evolution is obtained by solving equation (9) subject to the divergence constraint (19).

Following the algorithm mentioned in (Lin et al., 2020), our preliminary results showed that the solution contained spurious dislocation densities when the dislocation reactions are involved. Subsequent analysis revealed that the coupling of the slip systems by dislocation reactions induced significant numerical errors because of the divergence constraint imposed on total dislocation density tensor. Hence, to resolve this problem, an alternative method of enforcing the divergence constraint is introduced here based upon the concept of virtual dislocations.

## 3. Virtual dislocations

The idea of virtual dislocations in CDD framework originated from the fact that the dislocation lines are the boundary of slip areas and the physical dislocation line must be closed by a slip trace (El-Azab, 2000b). Recently, in the context of DDD simulations (Stricker et al., 2018) and (Po and Ghoniem, 2015), virtual dislocations were introduced as a non-physical entity with zero Burgers vector that bounds the glide area of dislocations involved in reactions such as cross slip, junction formation, or annihilation. From the perspective of the individual slip systems, virtual dislocations can then be considered as internal slip traces that, together with physical dislocation lines, provide closed boundaries of the areas swept by dislocations on their glide planes in the crystal interior. Borrowing this idea from the discrete setup, virtual density in CDD is defined as the non-physical density that quantifies the amount of physical dislocation density in CDD that is involved in the dislocation reaction. Since this density is non-physical, its contribution to plastic deformation of crystals is zero. Its contribution to the long-range stress in the crystal is also ensured to be zero by virtue of the fact the associated virtual dislocation density tensor at every point is zero.

For every slip system two vector densities are defined: a glide density $\boldsymbol{\rho}_\text{g}$ and a virtual density $\boldsymbol{\rho}_\text{v}$. The glide density is the physical dislocation density that distorts the crystal elastically and contributes to the plastic distortion. Correspondingly, two dislocation density tensors are defined: a glide dislocation density tensor $(\boldsymbol{\alpha}_\text{g})$ and a virtual dislocation density tensor $(\boldsymbol{\alpha}_\text{v})$. The virtual dislocation density tensor $\boldsymbol{\alpha}_\text{v}$ is defined as follows.

$$\boldsymbol{\alpha}_\text{v} = \sum_i \boldsymbol{\rho}_\text{v}^i \otimes \boldsymbol{b}^i . \tag{20}$$

The definition of the virtual dislocation density is explained by considering the two dislocation reactions of interest, namely, cross slip and glissile junctions. Fig.1(a) illustrates cross slip. In this process, a loop that originally belonged to slip system 1 is now spread onto both slip system 1 and 2. This configuration can be viewed as two loops, the first contained in slip system 1 and the second in 2, such that part of the two loops overlap at the intersection of the two cross slip planes. These parts are denoted by the dotted lines and they represent the virtual dislocation densities. The red dotted line in slip system 1 is the virtual density, $\boldsymbol{\rho}^1_{v_{cs},1\to 2}$, for slip system 1, and it corresponds to the closure of the glide density on slip system 1 corresponding to the part that has already cross slipped. The green dotted line in slip system 2 is the virtual density for slip system 2, $\boldsymbol{\rho}^2_{v_{cs},1\to 2}$ that closes the loop on slip system 2. The cross-slipped dislocation density from slip system 1 after bowing out is represented as $\boldsymbol{\rho}^2_g$. The rate of change of virtual densities of slip system 1 and 2 for a cross-slip reaction is given by equation below.

$$\dot{\boldsymbol{\rho}}^1_{v_{cs},1\to 2} = -\dot{\boldsymbol{\rho}}^2_{v_{cs},1\to 2} = \dot{r}_{i\to j}\boldsymbol{\rho}^1_g \tag{21}$$

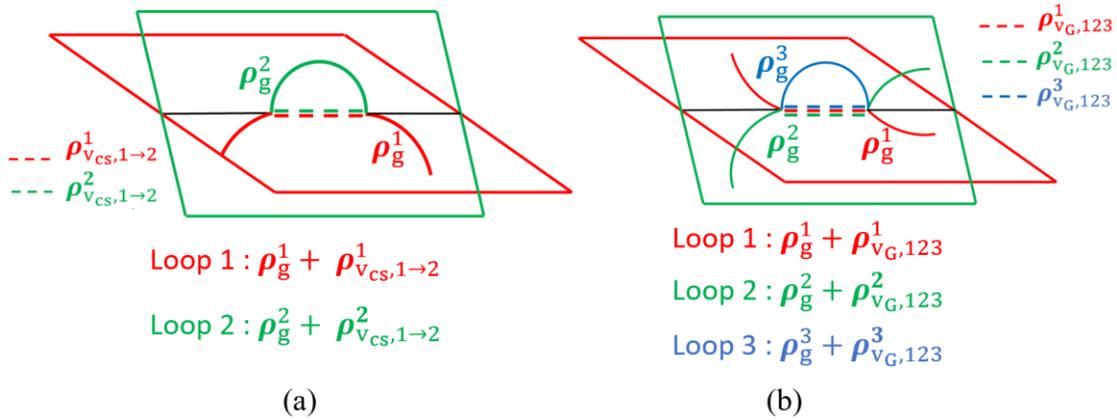

Fig. 1. Schematic showing the configuration of dislocations involved in (a) cross slip and (b) glissile junction reaction.

Since the rates of change of virtual dislocation densities of every pair of reacting slip systems are equal in magnitude but opposite in direction, their net contribution to the virtual dislocation density tensor amounts to zero. In mathematical terms, this is written in the form:

$$\boldsymbol{\alpha}_v = \boldsymbol{\rho}^1_{v_{cs},1\to 2}\otimes\boldsymbol{b}^1 + \boldsymbol{\rho}^2_{v_{cs},1\to 2}\otimes\boldsymbol{b}^2 = \left(\boldsymbol{\rho}^1_{v_{cs},1\to 2} - \boldsymbol{\rho}^1_{v_{cs},1\to 2}\right)\otimes\boldsymbol{b}^1 = 0 \tag{22}$$

Similarly, consider the case of glissile junction in Fig. 1(b). The glissile junction configuration can be thought of as three dislocation loops contained in their respective slip system such that a portion of each loop overlaps which is represented by the dotted lines. In this case, the red and green dotted lines correspond to virtual density in slip system 1 and 2, $\boldsymbol{\rho}^1_{v_G,123}$ and $\boldsymbol{\rho}^2_{v_G,123}$ respectively, which represents the part of the dislocation loop on slip system 1 and 2 involved in the glissile junction reaction and is no longer available to glide. The blue dotted line is the virtual density for slip system 3, $\boldsymbol{\rho}^3_{vG,123}$ that closes the loop on slip system 3. In Fig.1 (b), the dislocation bowing out after the glissile junction reaction between slip system 1 and 2 is represented as $\boldsymbol{\rho}^3_g$. The rate of change of virtual densities of slip system 1,2 and 3 for a glissile junction reaction is given by the equation below.

$$\dot{\boldsymbol{\rho}}^1_{vG,123} = \dot{\boldsymbol{\rho}}^2_{vG,123} = -\dot{\boldsymbol{\rho}}^3_{vG,123} = \dot{r}_{G,123}(\boldsymbol{\rho}^1_g \cdot \boldsymbol{e}_{12})(\boldsymbol{\rho}^2_g \cdot \boldsymbol{e}_{12})\boldsymbol{e}_{12} \tag{23}$$

The rate of change of virtual density on the third slip system $\dot{\boldsymbol{\rho}}^3_{vG,123}$ is equal in magnitude to $\dot{\boldsymbol{\rho}}^1_{vG,123}$, $\dot{\boldsymbol{\rho}}^2_{vG,123}$ but opposite in direction. Hence, the net contribution of these virtual densities to virtual dislocation density tensor can be shown to be zero using Frank's rule as follows

$$\boldsymbol{\alpha}_v = \boldsymbol{\rho}^1_{vG,123} \otimes \boldsymbol{b}^1 + \boldsymbol{\rho}^2_{vG,123} \otimes \boldsymbol{b}^2 + \boldsymbol{\rho}^3_{vG,123} \otimes \boldsymbol{b}^3 = \boldsymbol{\rho}^1_{vG,123}(\boldsymbol{b}^1 + \boldsymbol{b}^2 - \boldsymbol{b}^3) = 0, \tag{24}$$

when the reaction occurs such that $\boldsymbol{b}^1 + \boldsymbol{b}^2 = \boldsymbol{b}^3$. Generalizing the two cases discussed above to all types of dislocation reactions and slip systems, the virtual dislocation density evolution equation can be written as:

$$\dot{\boldsymbol{\rho}}^i_v = \dot{\boldsymbol{\rho}}^i_{vcs,i\to j} + \sum_j \dot{\boldsymbol{\rho}}_{vG,ijk} + \sum_j \dot{\boldsymbol{\rho}}_{vL,ij} + \sum_j \dot{\boldsymbol{\rho}}_{vH,ij} - \dot{\boldsymbol{\rho}}_{vcs,j\to i} \\ - \sum_{j,k} \dot{\boldsymbol{\rho}}_{vG,jki} - \dot{\boldsymbol{\rho}}_{vL^*,i} - \dot{\boldsymbol{\rho}}_{vH^*,i}. \tag{25}$$

where $\dot{\boldsymbol{\rho}}_{vL}$ and $\dot{\boldsymbol{\rho}}_{vH}$ correspond to the rate of change of virtual density due to the formation of Lomer and Hirth junctions, respectively, and $\dot{\boldsymbol{\rho}}_{vL^*}$ and $\dot{\boldsymbol{\rho}}_{vH^*}$ correspond to the rate of change of virtual density due to the unzipping of Lomer and Hirth junctions, respectively. Based on the above definitions, the dislocation configuration can be envisioned as an ensemble of loops represented with continuous densities, all made up of the sum of the glide and virtual dislocation densities entirely contained in the individual slip systems. From here onward, the sum of both the glide, $\dot{\boldsymbol{\rho}}^i_g$, and virtual, $\dot{\boldsymbol{\rho}}^i_v$, densities will be referred to as boundary vector density, $\boldsymbol{\rho}^i_{bd}$, which is the density characterizing the boundary of the slipped area on a given slip system. The evolution of this boundary vector density can be obtained by summing equation (18) and equation (25). Noting that the reaction terms in both the equations are equal and opposite, they drop out and we end up with the boundary vector density satisfying: $\dot{\boldsymbol{\rho}}^i_{bd} = \nabla \times (\boldsymbol{v}^i \times \boldsymbol{\rho}^i_{bd})$. By taking the divergence of this equation and noting that the divergence of the curl is zero, the boundary vector density will always be divergence free provided the initial configuration is divergence free. Hence, the dislocation closure can be ensured by enforcing the divergence-free condition on the boundary vector density on each slip system. That is,

$$\nabla \cdot \left(\boldsymbol{\rho}_{\text{bd}}^i\right) = \nabla \cdot \left(\boldsymbol{\rho}_{\text{g}}^i + \boldsymbol{\rho}_{\text{v}}^i\right) = 0. \tag{26}$$

The new form of the divergence constraint, equation (26), makes a significant enhancement in the accuracy of the numerical solution of the current CDD framework.

## 4. Numerical implementation

### 4.1 Time discretization

As mentioned in the earlier section, the CDD model consists of two parts: stress-equilibrium problem and the dislocation kinetic problem. The stress-equilibrium problem is solved in the same way as in (Lin et al., 2020). As such, only the numerical scheme to solve the kinetic equations is discussed in detail in this section. The dislocation kinetic equation describes two different physics, the transport and reactions of dislocations. The dislocation transport term in the kinetics equation is a curl type convection term describing the movement of dislocations under the local stress, whereas the dislocation reactions term describes the creation and removal of dislocations for various slip systems. Typically, multi-physics problems are harder to solve since the time scale associated with each physics and the solution method required to get a stable solution may vastly be different. A common technique used to overcome this hurdle is to solve the transport-reaction equation using the operator splitting, wherein is the transport and reaction physics are solved separately. In the current work, the dislocation transport part is solved first subjected to the divergence constraint to update the vector dislocation density. Then, the updated vector dislocation density is used to solve the dislocation reaction part. In doing so, the alteration of the densities due to the exchange between the glide and virtual densities does not influence the divergence-free condition (19) that is enforced during the transport solution step.

The numerical discretization for the transport part is written in the form:

$$\frac{\boldsymbol{\rho}_{\text{g}}^{i,t+1} - \boldsymbol{\rho}_{\text{g}}^{i,t}}{\Delta t} = \nabla \times \left(\boldsymbol{v}_{\text{g}}^{i,t} \times \boldsymbol{\rho}_{\text{g}}^{i,t+1}\right), \tag{27}$$

with $\boldsymbol{\rho}_{\text{g}}^{i,t+1}$ and $\boldsymbol{\rho}_{\text{g}}^{i,t}$ being the glide density at the beginning and end of the time step $\Delta t$. As indicated above, the velocity field $\boldsymbol{v}_{\text{g}}^{i,t}$ is supplied from the previous step due to the staggered scheme of solving the crystal mechanics and transport subproblems. Aside from that, the transport equations themselves are solved using an implicit scheme, which ensures numerical stability of the solution. The transport equation (27) is solved subject to the divergence constraint, which is now applied at the slip system level with the help of the virtual density,

$$\nabla \cdot \left(\boldsymbol{\rho}_{\text{g}}^{i,t+1} + \boldsymbol{\rho}_{\text{v}}^{i,t}\right) = 0. \tag{28}$$

Upon updating the dislocation density due to dislocation transport, the dislocation reactions are used to update the glide and virtual densities due to reactions. The discretized forms of the corresponding rate terms are:

Glissile junction reactions:

$$\left.\frac{\boldsymbol{\rho}_{\text{g}}^{i,t+1} - \boldsymbol{\rho}_{\text{g}}^{i,t}}{\Delta t}\right|_{\text{G}} = \sum_{j,k} \dot{r}_{\text{G},jki}(\boldsymbol{\rho}_{\text{g}}^{j,t+1} \cdot \boldsymbol{e}_{jk})(\boldsymbol{\rho}_{\text{g}}^{k,t+1} \cdot \boldsymbol{e}_{jk})\boldsymbol{e}_{jk} \tag{29}$$
$$- \sum_{j} \dot{r}_{\text{G},ijk}(\boldsymbol{\rho}_{\text{g}}^{i,t+1} \cdot \boldsymbol{e}_{ij})(\boldsymbol{\rho}_{\text{g}}^{j,t+1} \cdot \boldsymbol{e}_{ij})\boldsymbol{e}_{ij},$$

$$\left.\frac{\boldsymbol{\rho}_{\text{v}}^{i,t+1} - \boldsymbol{\rho}_{\text{v}}^{i,t}}{\Delta t}\right|_{\text{G}} = \sum_{j} \dot{r}_{\text{G},ijk}(\boldsymbol{\rho}_{\text{g}}^{i,t+1} \cdot \boldsymbol{e}_{ij})(\boldsymbol{\rho}_{\text{g}}^{j,t+1} \cdot \boldsymbol{e}_{ij})\boldsymbol{e}_{ij} \tag{30}$$
$$- \sum_{j,k} \dot{r}_{\text{G},jki}(\boldsymbol{\rho}_{\text{g}}^{j,t+1} \cdot \boldsymbol{e}_{jk})(\boldsymbol{\rho}_{\text{g}}^{k,t+1} \cdot \boldsymbol{e}_{jk})\boldsymbol{e}_{jk}.$$

Lomer junction reactions:

$$\left.\frac{\boldsymbol{\rho}_{\text{g}}^{i,t+1} - \boldsymbol{\rho}_{\text{g}}^{i,t}}{\Delta t}\right|_{\text{L}} = -\sum_{j} \dot{r}_{\text{L},ij}(\boldsymbol{\rho}_{\text{g}}^{i,t+1} \cdot \boldsymbol{e}_{ij})(\boldsymbol{\rho}_{\text{g}}^{j,t+1} \cdot \boldsymbol{e}_{ij})\boldsymbol{e}_{ij} + \dot{r}_{\text{L}^*,i}\boldsymbol{\rho}_{\text{L},i}^{t+1}, \tag{31}$$

$$\left.\frac{\boldsymbol{\rho}_{\text{v}}^{i,t+1} - \boldsymbol{\rho}_{\text{v}}^{i,t}}{\Delta t}\right|_{\text{L}} = \sum_{j} \dot{r}_{\text{L},ij}(\boldsymbol{\rho}_{\text{g}}^{i,t+1} \cdot \boldsymbol{e}_{ij})(\boldsymbol{\rho}_{\text{g}}^{j,t+1} \cdot \boldsymbol{e}_{ij})\boldsymbol{e}_{ij} - \dot{r}_{\text{L}^*,i}\boldsymbol{\rho}_{\text{L},i}^{t+1} \tag{32}$$

Hirth junction reactions:

$$\left.\frac{\boldsymbol{\rho}_{\text{g}}^{i,t+1} - \boldsymbol{\rho}_{\text{g}}^{i,t}}{\Delta t}\right|_{\text{H}} = -\sum_{j} \dot{r}_{\text{H},ij}(\boldsymbol{\rho}_{\text{g}}^{i,t+1} \cdot \boldsymbol{e}_{ij})(\boldsymbol{\rho}_{\text{g}}^{j,t+1} \cdot \boldsymbol{e}_{ij})\boldsymbol{e}_{ij} + \dot{r}_{\text{H}^*,i}\boldsymbol{\rho}_{\text{H},i}^{t+1}, \tag{33}$$

$$\left.\frac{\boldsymbol{\rho}_{\text{v}}^{i,t+1} - \boldsymbol{\rho}_{\text{v}}^{i,t}}{\Delta t}\right|_{\text{H}} = \sum_{j} \dot{r}_{\text{H},ij}(\boldsymbol{\rho}_{\text{g}}^{i,t+1} \cdot \boldsymbol{e}_{ij})(\boldsymbol{\rho}_{\text{g}}^{j,t+1} \cdot \boldsymbol{e}_{ij})\boldsymbol{e}_{ij} - \dot{r}_{\text{H}^*,i}\boldsymbol{\rho}_{\text{H},i}^{t+1} \tag{34}$$

Cross-slip:

$$\left.\frac{\rho_g^{i,t+1} - \rho_g^{i,t}}{\Delta t}\right|_{cs} = \dot{r}_{j \to i}\rho_g^{j,t+1} - \dot{r}_{i \to j}\rho_g^{i,t+1}, \tag{35}$$

$$\left.\frac{\rho_v^{i,t+1} - \rho_v^{i,t}}{\Delta t}\right|_{cs} = -\dot{r}_{j \to i}\rho_g^{j,t+1} + \dot{r}_{i \to j}\rho_g^{i,t+1}. \tag{36}$$

In the case of junction reactions, the discretized equations are non-linear, and are thus first linearized with the help of first order Taylor series expansion before using a linear solver. For example, consider the first term on the right side of equation (29). The term $\dot{r}_{G,jki}(\rho_g^{j,t+1} \cdot e_{jk})(\rho_g^{k,t+1} \cdot e_{jk})e_{jk}$ is of the form $f(x^{t+1}, y^{t+1})$ where $x$ and $y$ correspond to the dislocation densities in slip system $j$ and $k$. The first order multi-variate Taylor series expansion can be written as $f(x^{t+1}, y^{t+1}) = f(x^t, y^t) + f_x(x^t, y^t)(x^{t+1} - x^t) + f_y(x^t, y^t)(y^{t+1} - y^t)$, where $f_x$ and $f_y$ are the partial derivatives with respect to $x$ and $y$ respectively. Based on this, the first term can then be linearized as follows:

$$\begin{aligned}\dot{r}_{G,jki}(\rho_g^{j,t+1} \cdot e_{jk})(\rho_g^{k,t+1} \cdot e_{jk})e_{jk} = \dot{r}_{G,jki}\big[&(\rho_g^{j,t} \cdot e_{jk})e_{jk} \otimes e_{jk} \cdot \rho_g^{k,t+1} \\ &+ (\rho_g^{k,t} \cdot e_{jk})e_{jk} \otimes e_{jk} \cdot \rho_g^{j,t+1} \\ &- (\rho_g^{j,t} \cdot e_{jk})(\rho_g^{k,t} \cdot e_{jk})e_{jk}\big].\end{aligned} \tag{37}$$

Other junction reaction equations are linearized in the same fashion.

### 4.2 Finite element discretization of the dislocation transport equations

In this formulation, the glide dislocation density of each slip system is expressed as a 2D vector with components $\rho_{g1}$ and $\rho_{g2}$ within a local coordinate system in which the right-handed coordinates $x_1$, $x_2$ and $x_3$ are, respectively, along the Burgers vector (screw dislocation orientation), the edge dislocation direction, and the slip plane normal. The dislocation transport equations are then assembled in terms of the glide dislocation density for slip system $i$ as follows

$$\left[[A_t] + \Delta t * \left(-[A_0] - [A_1]\frac{\partial}{\partial x_1} - [A_2]\frac{\partial}{\partial x_2}\right)\right]\{\rho_g^i\}^{t+1} = [A_t]\{\rho_g^i\}^t. \tag{38}$$

Similarly, the divergence constraint can be assembled as

$$[A_{\text{div}}]\{\rho_g^i\}^{t+1} = -[A_{\text{div}}]\{\rho_v^i\}^t, \tag{39}$$

where $\{\rho_g^i\} = \{\rho_{g1}^i, \rho_{g2}^i\}^T$ corresponds to the components of glide dislocation density of slip system $i$ at time step $t$ and $\{\rho_v^i\} = \{\rho_{v1}^i, \rho_{v2}^i\}^T$ corresponds to the components of virtual dislocation density slip system $i$ at time step t and $[A_t]$, $[A_0]$, $[A_1]$, $[A_2]$ are the coefficients matrices. For a single slip system, the coefficient matrices $[A_t]$, $[A_0]$, $[A_1]$, $[A_2]$ and $[A_{\text{div}}]$ are defined as follows

$$[A_t] = \begin{bmatrix} 1 & 0 \\ 0 & 1 \end{bmatrix}, \quad [A_0] = \begin{bmatrix} -\dfrac{\partial v_2}{\partial x_2} & \dfrac{\partial v_1}{\partial x_2} \\ \dfrac{\partial v_2}{\partial x_1} & -\dfrac{\partial v_1}{\partial x_1} \end{bmatrix}, \quad [A_1] = \begin{bmatrix} 0 & 0 \\ -v_2 & v_1 \end{bmatrix}, \tag{40}$$

$$[A_2] = \begin{bmatrix} -v_2 & v_1 \\ 0 & 0 \end{bmatrix}, \quad [A_{\text{div}}] = \begin{bmatrix} \dfrac{\partial}{\partial x_1} & \dfrac{\partial}{\partial x_2} \end{bmatrix}.$$

Within the finite element framework, the glide dislocation density of the $i$th slip system over an element with $M$ nodes can be expressed in terms of shape functions and the nodal values of that element as follows:

$$\begin{Bmatrix} \rho_{g1}^i \\ \rho_{g2}^i \end{Bmatrix} = N \begin{Bmatrix} \rho_{g1}^{i,1} \\ \rho_{g2}^{i,1} \\ \rho_{g1}^{i,2} \\ \rho_{g2}^{i,2} \\ \vdots \\ \rho_{g1}^{i,M} \\ \rho_{g2}^{i,M} \end{Bmatrix}, \quad \dfrac{\partial}{\partial x_1} \begin{Bmatrix} \rho_{g1}^i \\ \rho_{g2}^i \end{Bmatrix} = B_1 \begin{Bmatrix} \rho_{g1}^{i,1} \\ \rho_{g2}^{i,1} \\ \rho_{g1}^{i,2} \\ \rho_{g2}^{i,2} \\ \vdots \\ \rho_{g1}^{i,M} \\ \rho_{g2}^{i,M} \end{Bmatrix}, \quad \dfrac{\partial}{\partial x_2} \begin{Bmatrix} \rho_{g1}^i \\ \rho_{g2}^i \end{Bmatrix} = B_2 \begin{Bmatrix} \rho_{g1}^{i,1} \\ \rho_{g2}^{i,1} \\ \rho_{g1}^{i,2} \\ \rho_{g2}^{i,2} \\ \vdots \\ \rho_{g1}^{i,M} \\ \rho_{g2}^{i,M} \end{Bmatrix}, \tag{41}$$

with $\rho_{g1}^{i,K}$ and $\rho_{g2}^{i,K}$, $K \in \{1, M\}$, being the value of glide dislocation density components at all nodes, and $[N]$, $[B_1]$ and $[B_2]$ are matrices containing the collection of elemental shape functions, $\{N\}$, and their derivative with respect to $x_1$ and $x_2$, respectively. These matrices are expressed in the form:

$$[N] = \begin{bmatrix} N_1 & 0 & N_2 & 0 & \cdots & N_M & 0 \\ 0 & N_1 & 0 & N_2 & \cdots & 0 & N_M \end{bmatrix} \tag{42}$$

$$[B_1] = \begin{bmatrix} \dfrac{\partial N_1}{\partial x_1} & 0 & \dfrac{\partial N_2}{\partial x_1} & \cdots & \dfrac{\partial N_M}{\partial x_1} & 0 \\ 0 & \dfrac{\partial N_1}{\partial x_1} & 0 & \cdots & 0 & \dfrac{\partial N_M}{\partial x_1} \end{bmatrix}; \tag{43}$$

$$[B_2] = \begin{bmatrix} \dfrac{\partial N_1}{\partial x_2} & 0 & \dfrac{\partial N_2}{\partial x_2} & \cdots & \dfrac{\partial N_M}{\partial x_2} & 0 \\ 0 & \dfrac{\partial N_1}{\partial x_2} & 0 & \cdots & 0 & \dfrac{\partial N_M}{\partial x_2} \end{bmatrix}.$$

With these expressions at hand, the dislocation transport equations (38) can be rewritten in terms of the nodal degrees of freedom in a more compact form,

$$[L]\{\rho_g^i\}^{t+1} = [W]\{\rho_g^i\}^t, \tag{44}$$

where the matrices $[L]$ and $[W]$ given by

$$[L] = [A_t][N] - \Delta t * ([A_0][N] + [A_1][B_1] + [A_2][B_2]), \tag{45}$$

$$[W] = [A_t][N]. \tag{46}$$

Likewise, the divergence constraint can be expressed in the form:

$$[B_{\text{div}}]\{\rho_g^i\}^{t+1} = -[B_{\text{div}}]\{\rho_v^i\}^t, \tag{47}$$

With

$$[B_{\text{div}}] = \begin{bmatrix} \frac{\partial N_1}{\partial x_1} & \frac{\partial N_1}{\partial x_2} & \frac{\partial N_2}{\partial x_1} & \frac{\partial N_2}{\partial x_1} & \cdots & \cdots & \frac{\partial N_M}{\partial x_1} & \frac{\partial N_M}{\partial x_2} \end{bmatrix}. \tag{48}$$

The above set of equations can be generalized to all slip systems and solved using Least Squares Finite Element Method (LSFEM) (Jiang, 2011; Xia and El-Azab, 2015; Lin et al., 2020). Skipping details, the global system of equations takes on the form:

$$[K]\{\rho_g\}^{t+1} = \{P\}, \tag{49}$$

with $[K]$ and $\{P\}$ given by

$$[K] = \int \left([L^T][L] + ch^2[B_d^T][B_d]\right) d\Omega$$
$$\{P\} = \int \left([L^T][W]\{\rho_g\}^t + ch^2[B_d^T][B_d]\{\rho_v\}^t\right) d\Omega. \tag{50}$$

In the above, the parameter $c$ corresponds to the control parameter used to impose the divergence constraint and $h$ is the mesh size.

Solving the system (49) will update the glide density due to dislocation transport on all slip systems in a decoupled fashion, meaning that equation (49) can be solved for one slip system at a time, thanks to the introduction of the virtual density. The dislocation reactions can then be used to update the glide and virtual densities.

## 5. Results

The significance of the virtual dislocation density is highlighted first by comparing the results for some test problems with and without consideration of the virtual density in the dislocation transport-reaction formulation in Section 5.1-5.3. Since the purpose of these numerical test problems is to demonstrate the effectiveness of the new divergence constraint, the system dynamics is evolved by enforcing a constant velocity. In addition, the rate coefficients for cross-slip and glissile junction formation are assumed, which does not impact on the qualitative nature of the results. Following this, the results for full bulk simulations are then presented in Section 5.4. In all the simulations, a crystal volume of dimensions 5µm×5µm×5.303µm was used, with periodic boundary conditions. The mesh used consists of tetrahedron and pyramid elements as proposed by (Xia and El-Azab, 2015) and the mesh size used is around 62.5nm on the edge. This value for mesh size was chosen so that the distance between two parallel planes on the mesh can be on the order of the annihilation distance of opposite mixed dislocations, thus conforming to the constraint of the bundle representation of dislocations; see also Xia et al., (2016). The initial dislocation configuration was created in the form of dislocation loop bundles which refers to a density-based representation of dislocations continuously distributed so as to have a single line orientation at every point. These loop bundles are gaussian distributed density over the bundle cross section, with the center of the dislocation loop bundle and the loop bundle radius and width as parameters. For brevity, from here onward, we will refer to the dislocation loop bundles as loops, the formulation without the virtual density as the *coupled formulation* and that with the virtual density as *decoupled formulation*. The reader is reminded again that, in the coupled formulation, the divergence constraint is enforced on all slip systems at once and therefore the scheme solves for all densities in a coupled fashion, versus enforcing the divergence constraint on individual slip systems in the decoupled formulation and thus solving for the slip system densities individually.

### 5.1 Dislocation loop expansion under a prescribed velocity

In this test, a single dislocation loop in a FCC crystal on the slip system 1, $(111)[0\bar{1}1]$, was allowed to expand under a constant applied velocity of 0.03 µm/ns normal to its line tangent everywhere. The loop expansion problem was solved by using the dislocation transport equation (28) coupled with a divergence constraint. In the case of a single loop, there is no dislocation reaction involved and hence the divergence constraint can be applied to the slip system mentioned above in a straightforward fashion. However, in order to show the effect of coupling of slip system solutions through the divergence constraint, this problem was also solved by coupling slip system 1 with its cross-slip system 2, $(\bar{1}11)[0\bar{1}1]$, (coupled formulation), with the latter having no dislocations initially. Ideally, the latter slip system should continue to have no density during the evolution. Fig. 2 shows the results of both formulations; the decoupled solution is shown in part (a) while the coupled solution is shown in part (b). Although cross slip was not activated, the coupled formulation populates dislocations on slip system 2 due to numerical errors that grow with time. The fact that coupling introduces spurious densities serves as a motivation to use the decoupled formulation, which is made possible by introducing the virtual dislocation density. As the simulation progresses, these spurious densities increase in magnitude, which can be seen in Fig. 3.

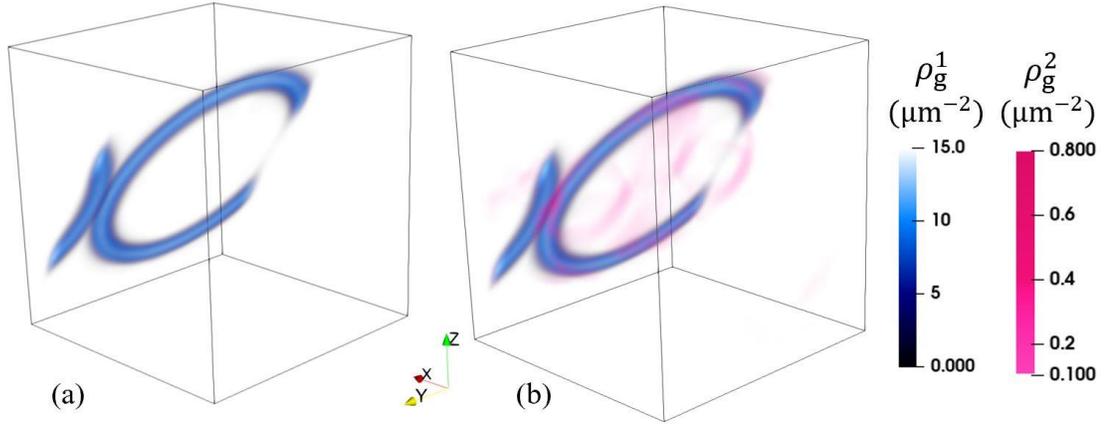

Fig. 2. Loop expansion test. A single loop was solved once using the decoupled formulation and another time using the coupled formulation. The results are shown in parts (a) and (b), respectively, after 28 ns (30-time steps).

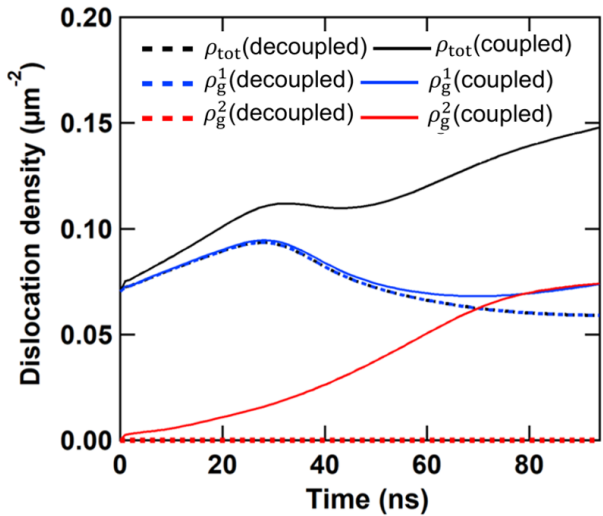

Fig. 3. Evolution of the average dislocation density in the dislocation loop expansion test. The dashed and solid lines correspond to decoupled and coupled formulations, respectively. The density $\rho_{tot}$ is the sum of $\rho_g^1$ and $\rho_g^2$. In the case of the decoupled formulation, the total density coincides with $\rho_g^1$ as should be.

## 5.2 Cross slip test with a prescribed applied velocity

Cross slip is a thermally activated process through which a dislocation changes its glide plane to a cross slip plane while retaining the same Burgers vector. In this test case, a single loop on the slip system 1, $(111)[0\bar{1}1]$, was considered. The simulation was setup such that cross-slip is allowed to happen for all dislocations with screw orientation. The velocity on the cross-slip plane was arbitrarily taken to be 0.1 µm/ns compared to 0.01 µm/ns on the original plane. This transport-reaction problem was solved using both coupled and decoupled formulations with a constant cross slip rate ($\dot{r}_{1\to 2} = 0.1$/ns). Fig. 4 (a) and (b) shows the results with the decoupled and coupled formulations, respectively after 8 ns (30 time steps). It is clearly shown that the coupled solution results in spurious density of dislocation on the cross-slip system. Such spurious densities result mainly through the divergence constraint. The effect of these spurious densities on the total density evolution in the domain of the simulation is shown in Fig. 5. While both solutions for the decoupled and coupled formulations on the main slip system are fairly close to each other, the solutions on the cross-slip plane for the two formulations differ significantly, thus resulting in a significant difference in the total density.

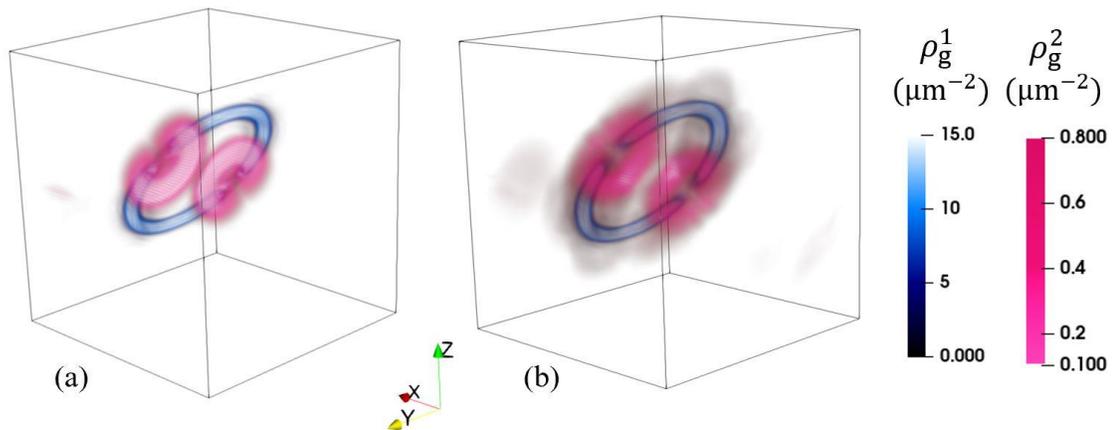

Fig. 4. Dislocation cross slip simulated using (a) the decoupled and (b) coupled formulations. In the case of decoupled formulation, the density on the cross-slip plane has no spurious values as it is the case in the coupled formulation.

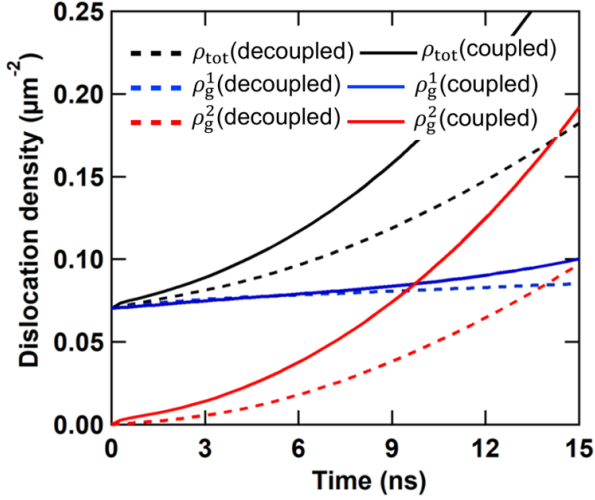

Fig. 5. Evolution of the average dislocation density for the cross-slip test with constant velocity on the glide plane and cross slip planes. The dashed and solid lines correspond to decoupled and coupled formulations, respectively. The density $\rho_{tot}$ is the sum of $\rho_g^1$ and $\rho_g^2$.

### 5.3 Glissile junction reaction under constant applied velocity

Glissile junction reactions represent a major mechanism of dislocation multiplication in multi-slip problems (Stricker et al., 2015; Stricker et al., 2018). In this test case, this dislocation reaction was studied by placing two dislocation loops on two slip systems, $(111)[0\bar{1}1]$ and $(\bar{1}11)[101]$. The dislocation loops on both slip systems were allowed to expand by imposing a constant velocity of 0.03 µm/ns. The glissile junction reaction between the dislocations was allowed to occur if the orientation criteria proposed by (Madec et al., 2002; Lin and El-Azab, 2020) was satisfied. This transport-reaction problem was solved with a constant glissile junction rate ($\dot{r}_{G,123} = 0.5$ µm$^2$/ns) using both coupled and decoupled formulations. The results after 28 ns (30-time steps) are shown in Fig. 6. From Fig. 6(b), it is evident that the coupled formulation populates slip system 3 with density at places where the glissile junction reaction is not happening. Such spurious density results from the coupling of the three slip systems through the divergence constraint. The effect of the spurious density is observed in the dislocation density evolution Fig. 7, where the total density $\rho_{tot}$ in the coupled formulation case is always higher.

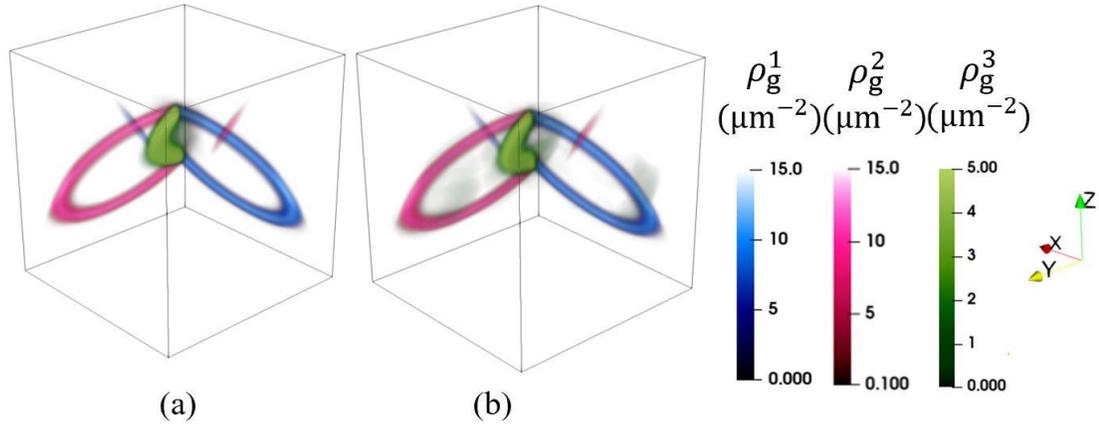

(a)              (b)

Fig. 6. The dislocation configuration for a glissile junction reaction from two loops simulated using the (a) decoupled and (b) coupled formulations. Again, spurious density was observed in the coupled formulation case. The pink and blue lines on the blue and pink loops in both (a) and (b) correspond to the portion of loops entering from the other side of the cube due to periodic boundary condition.

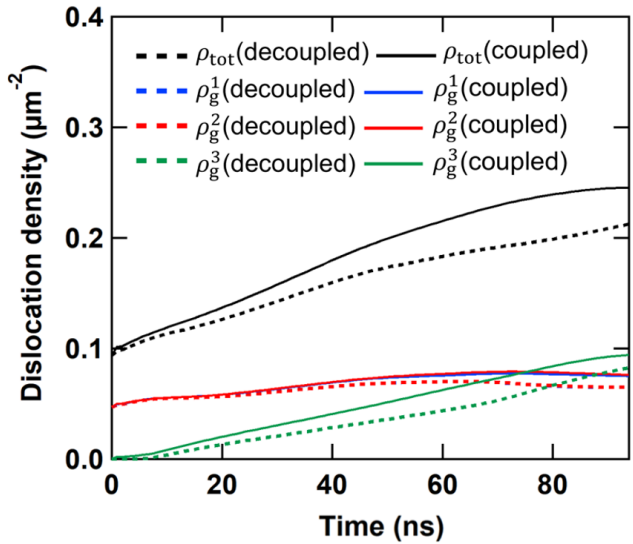

Fig. 7. Dislocation density evolution during a glissile junction reaction. Dashed and solid lines correspond to the decoupled and coupled formulations, respectively. The density $\rho_{tot}$ is the sum of $\rho_g^1$, $\rho_g^2$ and $\rho_g^3$.

## 5.4 Bulk Simulation

In this section, a simulation of a bulk FCC crystal under uniaxial loading for the multi-slip condition based on the decoupled formulation with virtual density is discussed. The material parameters used in this simulation are specified in Table 1. The system was subjected to a strain-rate controlled loading with an applied strain rate of 20/s along [001] direction. All 12 slip systems were populated with 10 circular dislocation loops each. The radius of the dislocation loop was chosen randomly between 2 µm and 4 µm and their centers were placed randomly in the crystal.

The dislocation velocity for each slip system $i$ is defined based on linear mobility law commonly used in literature (Zaiser et al., 2001; Yefimov et al., 2004; Lin and El-Azab, 2020) as follows

$$v^i = \frac{b}{B} \langle |\tau|^i - (\tau_0 + \tau_\text{p}^i) \rangle sgn(\tau^i) \tag{51}$$

where $b$ is the magnitude of the Burgers vector and $B$ is the drag coefficient. The $\langle \cdot \rangle$ corresponds to the Macauley bracket and $sgn$ function corresponds to the sign of the argument. $\tau^i$ is the resolved shear stress along slip system $i$ and $\tau_0$ is the lattice friction. $\tau_\text{p}^i$ is the Taylor hardening stress that accounts for short range interactions due to sessile junctions on slip system $i$ which is defined as

$$\tau_\text{p}^i = \alpha \mu b \sqrt{\sum_j a_{ij} \rho^j} \, f(\rho^i, \rho^f). \tag{52}$$

In the above, $a_{ij}$ is the average strength of an interaction between slip system $i$ and $j$, and $\rho^j$ is the dislocation density of the slip system $j$ interacting with slip system $i$ and $f(\rho^i, \rho^f)$ corresponds to a function that accounts for the dislocation pile up effect (Zhu et al., 2016).

Table 1. Bulk simulation parameters. The materials properties are those of a copper single crystal.

| Parameter | Value |
| --- | --- |
| Strain rate | 20s$^{-1}$ |
| Youngs Modulus | 112.5GPa |
| Poisson's ratio | 0.34 |
| Initial dislocation density | 1.5×10$^{12}$m$^{-2}$ |
| Burgers Vector | 0.25525nm |
| Drag coefficient | 5.5×10$^{-5}$Pas |

Although the virtual dislocation formalism in Section 3 incorporates all junction mechanisms, the sessile (Lomer and Hirth) junctions are not considered explicitly in the current bulk simulation test due to the lack a feasible implementation procedure for their unzipping within our CDD framework. Instead we choose to represent their effect indirectly using the Taylor hardening formula (52). In this formula, the coefficients corresponding to Lomer and Hirth lock interactions were obtained from (Madec and Kubin, 2017). The function $f(\rho^i, \rho^f)$ in equation (52) was introduced to prevent the Taylor hardening stress from increasing indefinitely with the density and, instead, force that terms to decay when the density reaches certain values, which is motivated by the work in (Zhu et al., 2016). Ideally, the explicit incorporation of formation and destruction of sessile junctions would prevent such an indefinite hardening. In our case, the form $f(\rho^i, \rho^f) = \left(1 + ae^{(b\rho^i/\rho^f - 1)}\right)^{-1}$ is considered, with *a* and *b* being parameters that determine the shape and the rate at which the function decays, and $\rho^f$ is the sum of the dislocation densities of the slip systems that react with slip system *i*.

Two simulations with two different decay function parameters were used to compare the impact of this function on the results. The corresponding simulations are named here CDD1 and CDD2. For CDD1, the values of *a* and *b* in the decay function are 0.00015 and 5, respectively. For CDD2, the values of *a* and *b* are 0.000008 and 13, respectively. The reaction rate coefficient for glissile junction formation was taken to be 0.5 $\mu m^2$/ns for both the simulations. The collinear annihilation rate was implemented based on (Lin and El-Azab, 2020) where the screw component of the smallest of the two reacting densities is fully annihilated. The contribution of collinear annihilation to the virtual dislocation density is added to that of cross slip because the two processes contribute to the same part of the virtual density. The cross-slip rate values were calibrated using a DDD simulation (Devincre et al., 2011) and transferred into the CDD model based on the procedure outlined in (Xia et al., 2016). The results obtained from the DDD model (Devincre et al., 2011) with the same initial conditions was used to establish a reference point for the CDD results.

The stress-strain curve and dislocation density curves of the CDD and the DDD simulations are shown in Figs. 8 and 9, respectively. The stress-strain curve of the CDD1 and CDD2 simulations follow a similar trend with a small difference in yield point as can be seen from Fig. 8(a). The difference in the decay function parameters affects the flow stress, which consequently affects the yield point and dislocation density evolution as observed in Fig. 8(a) and (b). The hardening rate of the CDD simulations varies in tandem with the dislocation density curve, which is evident by comparing Fig. 8(a) and (b), with the hardening rates being smaller than the DDD result in this case. It is to be noted that the microMegas model (Devincre et al., 2011) does not include glissile junction reactions, the presence of which is known to soften the material response at least initially. Fig. 9(a) and (b) shows the dislocation density evolution for all slip systems for both the CDD and DDD simulations, respectively. The dislocation densities on the eight active slip systems increase faster than those of the four inactive slip systems in both cases. However, the increase of dislocation density in the inactive slip systems is more pronounced in the case of CDD due to glissile junction formation, which couple the active and inactive slip systems and promote dislocation storage on the inactive slip systems. In the case of DDD simulation, the increase in density of slip system 8 is much higher compared to other active slip systems. This behavior resulted from the initial configuration setup and material parameters.

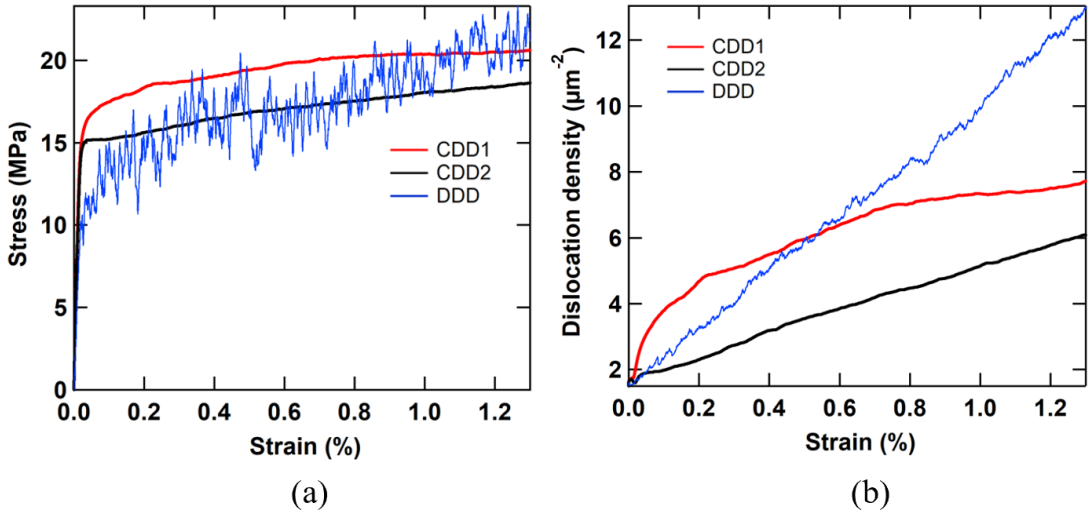

(a) (b)

Fig. 8. Comparison of CDD1, CDD2 and DDD simulation results. (a) Stress-strain curves. (b) Dislocation density evolution.

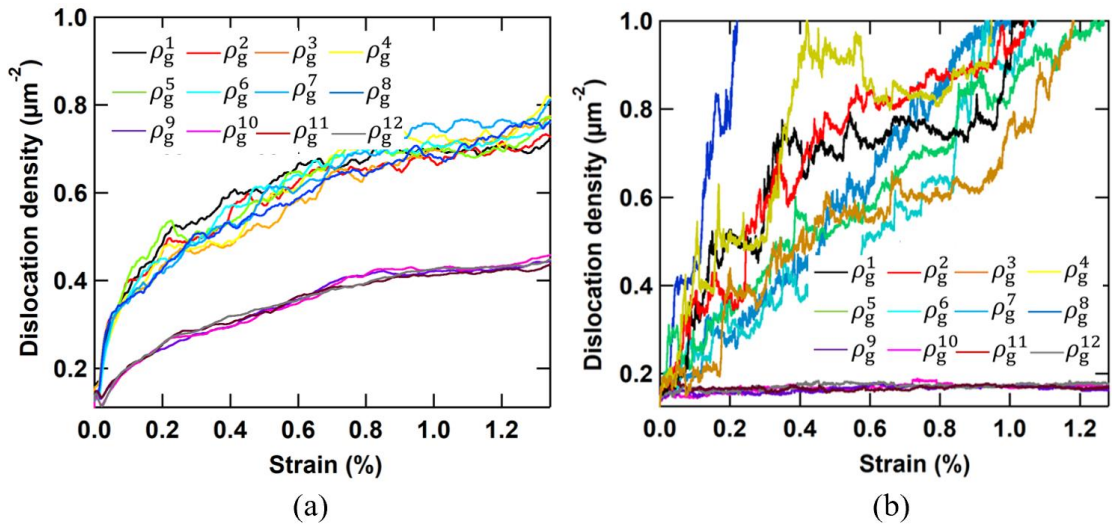

(a) (b)

Fig. 9. Evolution of the slip system dislocation densities for (a) CDD1 and (b) DDD simulations. The four nominally inactive slip systems for the [001] loading store more dislocations in the case of CDD1 due to glissile junction reactions.

### 5.4.1 Dislocation microstructure

The dislocation microstructure is visualized by plotting the scalar dislocation density field in the simulation domain. This density field is obtained by summing the magnitude of the vector dislocation densities of all slip systems,

$$\rho_{\text{scalar}} = \sum_{i=1}^{12} \|\boldsymbol{\rho}_g^i\|. \tag{53}$$

Fig. 10(a) shows the dislocation distribution (pattern) over the simulation domain from the CDD1 simulation at 0.6% strain. The pattern shows features along $<110>$ type directions, which lie along the intersections of the FCC slip planes and are Burgers vectors directions. Accumulation of dislocations along these directions is consistent with the fact that the Taylor hardening mechanism operates mostly at the interactions of different slip planes. Glissile junctions also form along those directions. Fig. 10(b) shows the virtual dislocation density at a strain of 0.6%, which is determined by summing the magnitude of virtual dislocation densities of all slip systems,

$$\rho_{\text{virtual}} = \sum_{i=1}^{12} \|\boldsymbol{\rho}_v^i\|. \tag{54}$$

The pattern in Fig. 10(b) also shows clear features along $<110>$ type directions. Since, the virtual dislocations serve as a mark of the crystal spots where dislocation reactions and cross slip took place, it is reasonable to assume that they also can demarcate density features along $<110>$ type directions. Typically, the dislocations that cross-slipped or resulted from junctions, the glissile in particular, depart from the spots where such events take place via glide. Nevertheless, the dislocations involved in the most recent junction reactions and cross slip may still be close to where these events took place and thus close to where the virtual dislocations accumulate which can be observed at certain locations by comparing Fig. 10(a) and (b).

Fig. 11. shows the dislocation microstructure over the $(\bar{1}11)$, $(111)$, $(010)$ and $(100)$ planes at a strain value of 0.6%. The corresponding virtual density patterns over the same planes and the same strain are shown in Fig. 12. In all these planes, it is noticed again that both the dislocation pattern and virtual dislocation pattern show higher density along the $<110>$ type directions. The areas of high dislocation density correlate with the high virtual density areas at certain locations similar to Fig. 10. This indicates that the dislocation pattern at any moment during the evolution of the system, although primarily caused by Taylor hardening mechanism, also depends on dislocation cross slip and junction reactions.

Fig. 10. Dislocation microstructure from the CDD1 simulation at 0.6% strain. (a) Scalar dislocation density. (b) The virtual dislocation density.

Fig.11. Dislocation microstructure from the CDD1 simulation at 0.6% strain viewed on the (111), ($\bar{1}$11), (010), and (100) planes in (a) through (d), respectively.

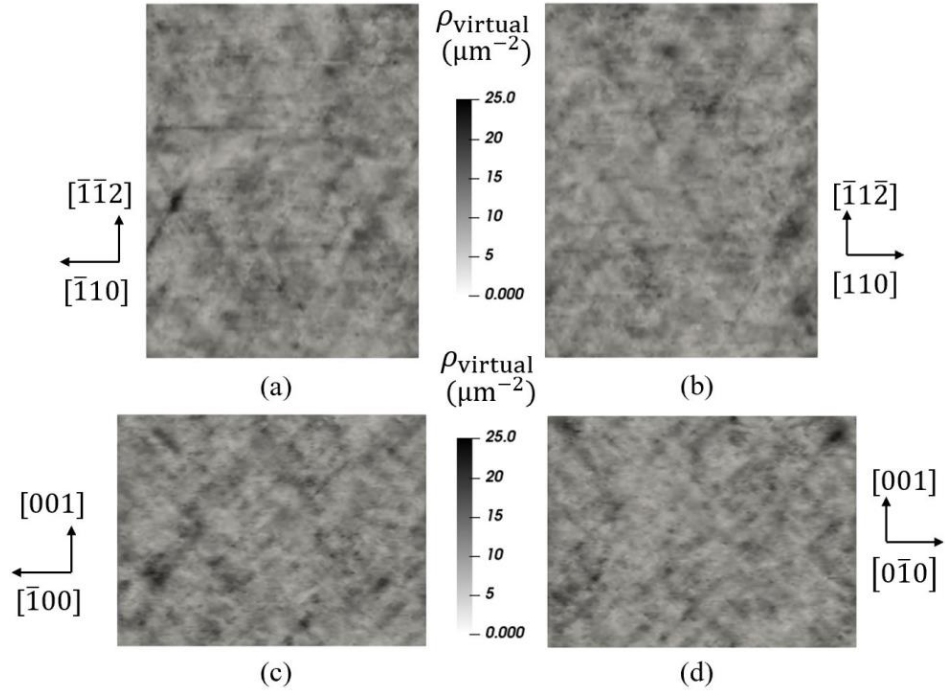

Fig. 12. Virtual dislocation pattern from the simulation CDD1 at 0.6% strain viewed on the (111), ($\bar{1}$11), (010), and (100) planes in (a) through (d), respectively.

### 5.4.2 Geometrically necessary dislocation density

The dislocation microstructure can also be characterized based on the pattern formed by geometrically necessary dislocations (GNDs). GNDs are represented as tensors which are calculated based on equation (7). The trace norm of the GND tensor ($\|\alpha_{\text{total}}\|$) is used as a measure to visualize the pattern formed by GNDs. Fig. 13 shows the GND pattern formed over different planes, with data coming from CDD1 simulation at 0.6% strain. The patterns on all planes show a strong feature along the $<110>$ type directions, which is consistent with the results shown in Figs. 11 for the scalar dislocation density. The GND distribution carries the net Burgers vector distribution within the deformed grains and is thus important in fixing both the lattice rotation and the elastic strain fields, both of which are the subject of modern 3D synchrotron X-ray microscopy (Larson et al., 2002; Levine et al., 2016; Poulsen et al., 2003). These fields are also essential in understanding a phenomenon like recrystallization.

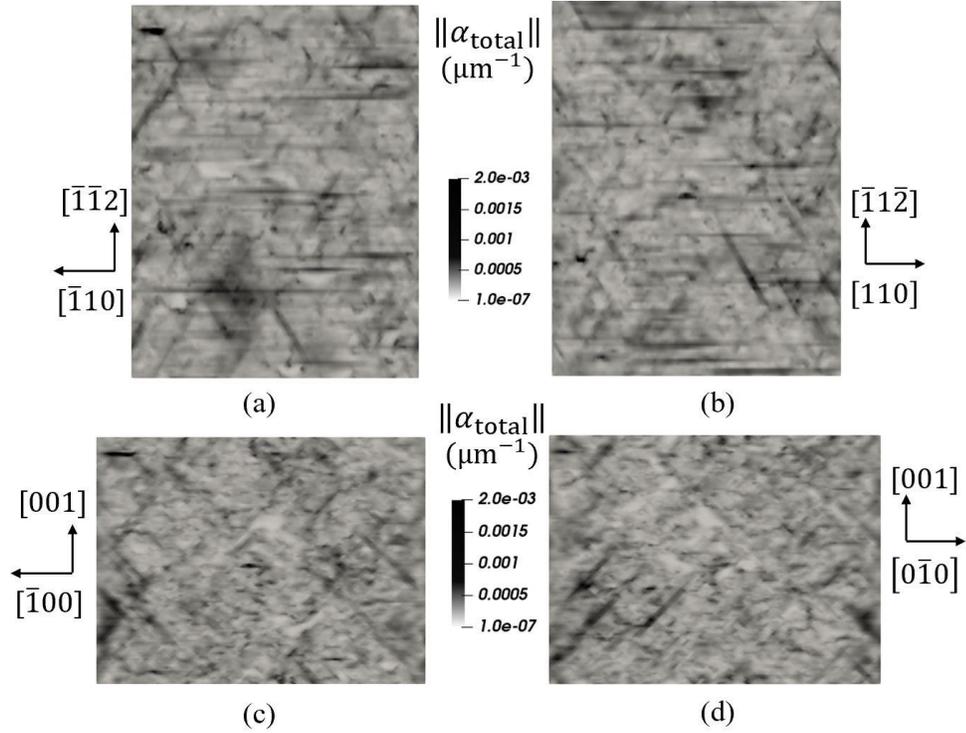

Fig. 13. GND pattern from the simulation CDD1 computed as the trace norm of the dislocation density tensor at 0.6% strain, and viewed on the (111), ($\bar{1}$11), (010), and (100) planes in (a) through (d), respectively.

### 5.4.3 Lattice rotation

The lattice rotation vector, $\Omega_k$, is expressed in terms the rotation tensor, $\omega_{ij}$,

$$\Omega_k = \frac{1}{2} \varepsilon_{ijk} \omega_{ij} \tag{55}$$

The lattice rotation tensor itself is the anti-symmetric part of the elastic distortion tensor. Fig. 14 shows the lattice rotation vector represented as an RGB plot wherein the purely red, green and blue colors correspond to the $x$, $y$ and $z$ component of the lattice rotation vector. The RGB plot is obtained by normalizing the lattice rotation vector and mapping the scalar value of each component to RGB values. This field provides information about the abrupt lattice direction change (misorientation) across space, which demarcates the subgrain structure in deformed metals. It can provide the so-called geometrically necessary boundaries (GNBs) and incidental dislocation boundaries (IDBs) observed in TEM experiments (Godfrey and Hughes, 1999; Hughes et al., 1997, 2003). This subgrain structure is also directly measured in 3D using synchrotron X-ray microscopy (Larson et al., 2002; Levine et al., 2016; Poulsen et al., 2003). The subgrain structure and the associated GND fields are both important in the context of recrystallization (Humphreys and Hatherly, 2012).

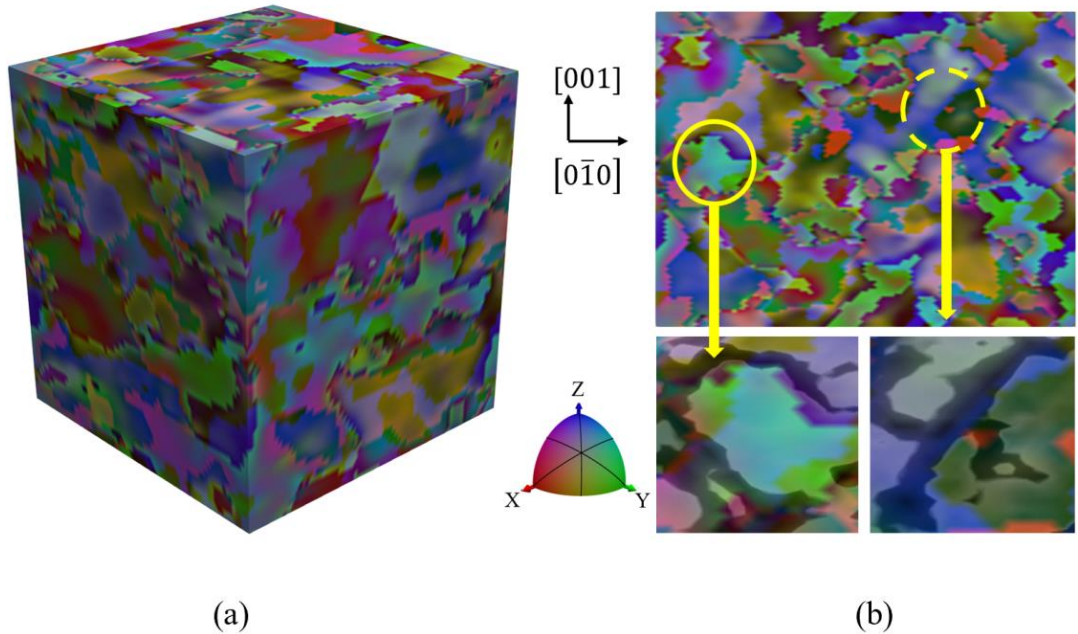

(a) (b)

Fig.14. Lattice rotation shown as RGB plot where red, green and blue corresponds to the lattice rotation components $\Omega_1$, $\Omega_2$ and $\Omega_3$ while the rest of the colors represent the intermediate components of the lattice rotation vector at 0.6% strain. (a) Bulk (b) slice along (100) plane. Two regions were enlarged to show the GND density distribution (translucent black lines) superimposed on the lattice rotation fields.

Regions with no discernible change in orientation, mild change in orientation and drastic change in orientation can be observed in Fig.14. These features can be used to distinguish the dislocation density distribution surrounding these regions. The uniform lattice orientation in a region corresponds to a lack of plastic strain gradient within that region, which, indicates that the region is free of dislocations. This type of profile can be observed in the region marked by the solid yellow circle where the lattice rotation is predominantly oriented along the direction corresponding the color light blue in the RGB plot. On the other hand, drastic change in the lattice orientation corresponds to the presence of high strain gradient, which in turn indicates geometrically necessary dislocation density accumulation in that region. This type of profile can be observed in the region marked by the dashed yellow circle where the lattice rotation changes from dark green, which corresponds to a direction close to $y$, to blue and bluish green which corresponds to a direction close to $z$ direction. This can be verified by overlaying the GND density distribution on the lattice rotation distribution for the regions discussed. The snippets of the zoomed in regions show the GND density superimposed on the lattice rotation fields. As expected inside the solid yellow circle region, there is no GND density in the regions with uniform lattice rotation field. Also, in the dashed yellow circle region, there is a translucent black line separating the green and bluish region indicating the dislocation accumulation on the boundary. This shows that lattice rotation is inter-related with the GND density distribution.

While Fig. 14 shows the salient features of the lattice rotation field in 3D, these features can be collectively understood by looking at their statistics. Fig. 15 shows the probability distribution function of each individual component of lattice rotation field, fit to a Gaussian curve. The Gaussian curves were fit to the raw data based on the maximum likelihood method (MLE) using MATLAB software. As expected from the data, the distribution of $\Omega_3$ has a sharp peak and narrow width compared to the other two components. All the three components are distributed nearly symmetrically about the mean with a minor skewness which is similar to the trend observed in DDD simulations (Mohamed et al., 2015). Such a near-symmetrical feature of the distribution is characteristic of statistically homogeneous bulk plastic deformation. The mean values of the three lattice rotation components $\Omega_1$, $\Omega_2$, and $\Omega_3$ are $-3.7934 \times 10^{-6 \circ}$, $-5.1345 \times 10^{-6 \circ}$ and $-5.697 \times 10^{-6 \circ}$, respectively, which are all 0° for practical purposes, and their standard deviations are 0.0035°, 0.0036° and 0.0014°, respectively. The average lattice rotation for all three components is thus considered zero. The standard deviation of the two lattice rotation components $\Omega_1$ and $\Omega_2$ is more than twice that of the lattice rotation component $\Omega_3$. The full width at half maximum (FWHM) for the three components is 0.0082°, 0.0085° and 0.0033°, respectively. These statistics indicate that the lattice rotation components $\Omega_1$, $\Omega_2$, and $\Omega_3$ are in general different, and such differences might depend upon the initial dislocation density and its evolution, but definitely upon loading direction.

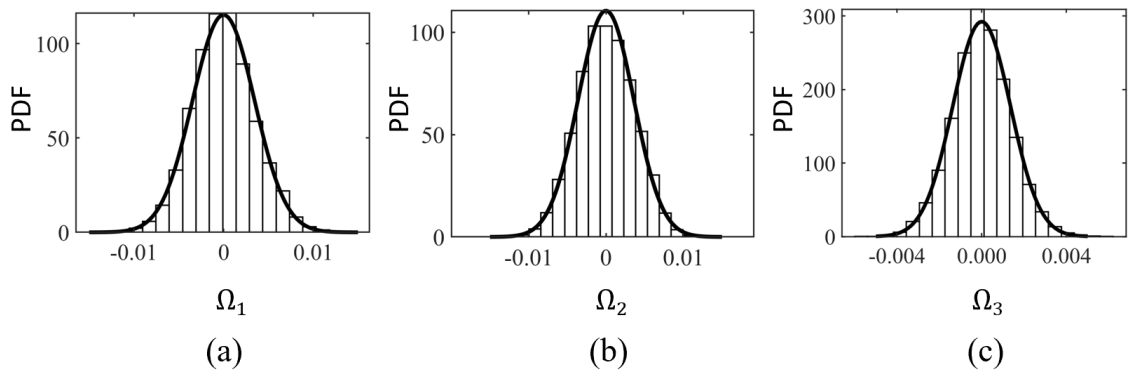

Fig.15. Probability distribution function of the three lattice rotation components in (°) at 0.6% strain. The panels (a), (b) and (c) display the probability distribution functions of $\Omega_1$, $\Omega_2$ and $\Omega_3$, respectively. Gaussian curves (red) are fit onto the raw data of the individual lattice rotation

component distributions based on maximum likelihood estimation (MLE) using MATLAB software.

## 6. Discussion

The results discussed in the previous sections have showcased the significance of the concept of virtual dislocation density; namely that, the introduction of the virtual density enabled us to solve the dislocation density evolution equation on each slip system individually as opposed to solving them in a coupled fashion. Achieving decoupling of the density evolution equations during the numerical solution was also enabled by the use of the operator splitting scheme in solving the transport and the reaction parts, while treating the virtual density in the divergence constraint in a semi-implicit fashion. It is important to note that solving the density evolution equation numerically in the decoupled formulation offers three advantages. The first is lowering the computational cost to solve the linear system of equations for slip system activity individually in comparison to solving all slip systems together. The second is the control of the accuracy of the solution and suppressing the spurious errors arising from the coupled divergence constraint. The third is that keeping track of the density of the dislocations involved in reactions via the virtual density on every slip system makes it possible to model processes like unzipping of sessile dislocation junctions, a possible future extension of the current work.

The reader must be alerted to the fact that the current work is but a step toward completing the CDD framework. Until that point is reached, the comparison of CDD results with DDD results for the same simulation conditions will always show some differences. As far as bulk simulations are concerned, such differences in results can be explained in terms of the physics built into the two types of models. One major difference between the two sets of simulations is the lack of glissile junction mechanism in the simulations based upon the microMegas DDD model used here (Devincre et al., 2011). The incorporation of glissile junction reactions to CDD equations couples the density evolution of multiple slip systems together. Consequently, the dislocation density increase in a slip system no longer solely depends only on the plastic slip happening in the slip system but rather on the activity of the reacting systems as well. It is evident from Fig. 9(a) that dislocation density increase within inactive slip systems is due to glissile junction reactions – a mechanism that contradicts the classical interpretation of dislocation storage during plastic deformation, which suggests that the dislocation storage on a given slip system is mainly due to the slip activity within that slip system. The dislocation density in active slip systems increases with the applied strain thus increasing the possibility of glissile junction reactions with dislocations within the inactive systems. This results in the steady increase in dislocation density of inactive slip systems as observed in Fig. 9(a). On the other hand, due to the lack of glissile junction mechanism, there is no increase in the density of inactive slip systems in the DDD simulation as can be seen from Fig. 9(b). Although the cross-slip mechanism also aids in transferring slip activity and hence dislocations between slip systems, there is still no net increase in the dislocation density on the inactive slip systems in the DDD simulation conducted here because cross slip to the inactive slip systems is not operative.

Fig. 8(a) shows the stress-strain curves of CDD and DDD simulation. The CDD stress-strain curves have a slightly higher yield point and lower hardening rate compared to the DDD simulation. The yield point is sensitive to the initial dislocation configuration, and the CDD and DDD yield points can be calibrated to coincide with one another. However, the difference in the hardening rate can be attributed to the absence of an explicit representation of sessile junction formation mechanism in the current CDD simulations. The incorporation of sessile junctions will reduce the gliding dislocation density and therefore increase the hardening rate. Sessile junctions were not implemented in the current simulations due to the lack of a procedure for unzipping such junctions. Another possibility for the lower hardening rate obtained in the CDD simulation is the use of the decay function $f(\rho^i, \rho^f)$ in the Taylor hardening formula (see equation (52)). As discussed earlier, this function was used to emulate the destruction of sessile junctions and prevent indefinite accumulation of dislocations against each other in regions of high dislocation density, by choosing it to be a smoothly decaying step function in the density. Currently, there is no quantitative basis at this point to calibrate such function. Instead, its impact on the results was checked by varying the parameters $a$ and $b$ used in that function to correspond to the CDD1 and CDD2 simulations. The results in Fig. 8(a) show that the simulation with slower rate of decay (CDD1) has higher yield point compared to the simulation with faster rate of decay (CDD2). Consequently, the difference in flow stress also affects the dislocation density evolution between the two sets of simulations as can be seen from Fig.8(b). In all cases, the breakaway of dislocations upon reaching the decay threshold is of course expected to lead to rapid dislocation motion and possible over-annihilation. This is perhaps the reason the dislocation density in the CDD case rises at a slower rate than in the DDD counterpart, and hence the lower hardening rate in the case of CDD.

Ideally speaking, the Taylor hardening term should be replaced with a proper implementation of the correlation stress in 3D, and the hard junctions should be explicitly implemented. Two-dimensional CDD models showed that the correlation leads to multiple resistive stresses, some of which are Taylor-like (Groma et al., 2016; Wu et al., 2018). Preliminary results on the dislocation-dislocation correlation in 3D have been recently obtained (Anderson and El-Azab, 2020) and its implementation in CDD is underway.

# 7. Summary and conclusions

A novel formulation to incorporate dislocation reactions into vector density-based CDD models was proposed using the concept of virtual dislocation density. The latter was introduced to keep track of the dislocations involved in dislocation networks among various slip systems via cross-slip and junction reactions. Using the virtual density on each slip system as a complement to the glide density, it was easily possible to decouple the numerical solution of the transport-reaction equations of the dislocation density on all slip systems, thus expediting the solution and controlling the numerical errors associated with the enforcement of the divergence constraint. This new formulation was implemented based on the operator splitting approach, where the dislocation transport equation was solved first using the least-squares approach, with the divergence constraint implemented in a semi-implicit fashion.

The effectiveness of the new formulation was studied using a number of test cases. The results showed that the new formulation was able to enforce the dislocation continuity rigorously by preventing the accumulation of spurious dislocation densities observed in the coupled formulation. The significance of this was further emphasized by showing how the numerical errors snowballed as the simulation progresses through the dislocation density evolution graphs. The new formulation was then used to study the behavior of FCC crystals under monotonic loading. The stress-strain curves obtained from the simulations exhibited similar behavior as those obtained by DDD simulations, with some differences explained in terms of the different physics contained in the CDD and DDD simulations. As expected, the microstructural features obtained from the simulation results showed high density features along the $<110>$ type directions. Analysis of the virtual dislocation density microstructure revealed that the dislocation reactions played an important role in the generation of these microstructural features. Lattice rotation results revealed clearly the subgrain structure associated with the dislocation patterns obtained from the solution. The probability distribution functions of the lattice rotation components were also shown to agree with the DDD simulations results reported in (Mohamed et al., 2015).

This virtual density approach can be considered as a step forward in enabling the CDD approach to accurately capture the dislocation microstructure. Further improvements of this approach by explicitly incorporating sessile junction formation and removing the phenomenological Taylor hardening part of the model is in progress. In this case, the density information captured by the virtual dislocation density can be used to account for the unzipping of sessile junctions based on the local stress state. Furthermore, incorporating the information about dislocation correlations can help in improving the accuracy of the dislocation density transfer during the dislocation reaction process.

## Acknowledgements

This work is supported by the US Department of Energy, Office of Science, Division of Materials Sciences and Engineering, through award number DE-SC0017718 at Purdue University.

Xia, S., El-Azab, A., 2015. Computational modelling of mesoscale dislocation patterning and plastic deformation of single crystals. Modelling and Simulation in Materials Science and Engineering 23, 055009. https://doi.org/10.1088/0965-0393/23/5/055009

Yefimov, S., Groma, I., van der Giessen, E., 2004. A comparison of a statistical-mechanics based plasticity model with discrete dislocation plasticity calculations. Journal of the Mechanics and Physics of Solids 52, 279–300. https://doi.org/10.1016/S0022-5096(03)00094-2

Zaiser, M., Miguel, M.-C., Groma, I., 2001. Statistical dynamics of dislocation systems: The influence of dislocation-dislocation correlations. Physical Review B 64. https://doi.org/10.1103/PhysRevB.64.224102

Zhu, Y., Xiang, Y., Schulz, K., 2016. The role of dislocation pile-up in flow stress determination and strain hardening. Scripta Materialia 116, 53–56. https://doi.org/10.1016/j.scriptamat.2016.01.025
Page | 39